\documentclass[twocolumn]{aastex6}

\epsscale{1.17} 

\usepackage{epsfig}
\usepackage{amsmath}
\usepackage{amssymb}

\newcommand{\Msolar}{\mbox{$\rm M_{\odot}$}} 
\newcommand{\Rsolar}{\rm{R_{\odot}}} 
\newcommand{\Lsolar}{\rm{L_{\odot}}} 
\newcommand{\nHe}{n_{\rm He}} 
\newcommand{\iso}[2]{\mbox{$^{#1}{\rm #2}$}} 

\newcommand{\MWD}{M_{\rm{WD}}}
\newcommand{\MMS}{M_{\rm{MS}}}

\shorttitle{HeWD+MS mergers}
\shortauthors{Zhang et al.}

\begin{document}
\title{Evolution models of helium white dwarf--main sequence star merger remnants}

\author{Xianfei Zhang}
\affil{Department of Astronomy, Beijing Normal University, Beijing, 100875, China}
\email{zxf@bnu.edu.cn}

\author{Philip D.~Hall}
\affil{Armagh Observatory, College Hill, Armagh BT61 9DG, UK}

\author{C.~Simon Jeffery}
\affil{Armagh Observatory, College Hill, Armagh BT61 9DG, UK;\\
School of Physics, Trinity College Dublin, Dublin 2, Ireland}

\and

\author{Shaolan Bi}
\affil{Department of Astronomy, Beijing Normal University, Beijing, 100875, China}


\begin{abstract}
It is predicted that orbital decay by gravitational-wave radiation and tidal interaction will cause some close-binary stars to merge within a Hubble time.
The merger of a helium-core white dwarf with a main-sequence star can produce a red giant branch star that has a low-mass hydrogen envelope when helium is ignited and thus become a hot subdwarf.
Because detailed calculations have not been made, we compute post-merger models with a stellar evolution code.
We find the evolutionary paths available to merger remnants and find the pre-merger conditions that lead to the formation of hot subdwarfs.
We find that some such mergers result in the formation of stars with intermediate helium-rich surfaces.
These stars later develop helium-poor surfaces owing to diffusion.
Combining our results with a model population and comparing to observed stars, we find that some observed intermediate helium-rich hot subdwarfs can be explained as the remnants of the mergers of helium-core white dwarfs with low-mass main-sequence stars.
\end{abstract}


\section{Introduction}
Hot subdwarfs are extreme horizontal branch (EHB) stars located close to the helium main sequence in the Hertzsprung--Russell (HR) diagram.
Spectroscopically, they can be roughly divided into three classes:
subdwarf B (sdB), subdwarf O (sdO), and subdwarf OB (sdOB).
Most of these stars are thought to have helium-burning cores
and extremely low-mass ($<0.02\,\Msolar$) hydrogen-rich envelopes
\citep[surface helium number fraction $\nHe<1\%$;][]{Heber09}.
However, about $10\%$ of hot subdwarfs have He-strong-lined spectra,
and are known as helium-rich hot subdwarfs.
These He-rich subdwarfs can be further subdivided spectroscopically into
three groups with members showing, respectively, strong carbon lines  (C-type),
strong nitrogen lines (N-type) and both (CN-type). The identification of
formation channels for each of these groups offers a challenge to
the theory of stellar evolution.

Most of the helium-rich hot subdwarfs have a nearly pure helium surface with surface helium number fraction $\nHe>90\%$.
A small number of hot subdwarfs have $\nHe = 10$--$90\%$ and are referred
to as intermediate helium-rich (iHe-rich) hot subdwarfs.
A few of these stars have extraordinary surface compositions, with abundances of lead, zirconium, strontium and yttrium up to $10\,000$ times the solar value \citep{Naslim11,Naslim13}.

A possible channel to the formation of He-rich hot subdwarfs is the merger of two helium white dwarfs \citep[HeWDs;][]{Webbink84}.
Several close detached HeWD+HeWD binary systems have been observed \citep{Maxted99,Maxted2000}.
The orbital energy of such systems can be removed by the emission of  gravitational wave radiation, leading the orbit to decay and the stellar components, ultimately, to come into contact.
Sufficiently short-period HeWD+HeWD binaries are thus expected to merge within a Hubble time and become He-rich subdwarfs \citep{Saio00, Han02, Han03}.
The merger process itself is hot, leading to prompt nucleosynthesis,
and the subsequent evolution includes epochs of strong convection,
 both flash-driven and opacity-driven, leading to the exposure of nuclear products at the surface.
\citet{Zhang12a} argued that this mixing is sensitive to the overall merger mass
and that there is a correlation between mass and surface carbon abundance.
The predicted atmospheric abundances of the merger products are found to match those observed.

According to previous HeWD+HeWD merger calculations \citep{Zhang12a},
any remaining hydrogen on the surface layer of a HeWD is expected to be
converted to helium during the extremely hot initial phase of the merger.
Consequently, the merger remnants all have nearly pure helium surfaces.
Thus, HeWD+HeWD mergers may not produce hot subdwarfs with a substantial surface fraction of hydrogen.
The origin of iHe-rich hot subdwarfs and their unusual surface
abundances of exotic elements is therefore a puzzle for stellar evolution.

Another possible channel to the formation of hot subdwarfs is the merger of a HeWD with a low-mass main-sequence (MS) companion \citep{Clausen2011}.
In addition to the HeWD+HeWD white dwarf binaries, there are many short-period detached binary systems composed of HeWDs with main-sequence (MS) companions \citep{Zorotovic2011}.
For instance, \mbox{SDSS\,J121010.1+334722.9} is a cool $0.4\,\Msolar$ HeWD with a $0.16\,\Msolar$ M dwarf companion in a 3 hours eclipsing binary which probably formed through common envelope (CE) evolution \citep{Pyrzas12}.
In such systems, owing to gravitational-wave radiation, tidal interaction and magnetic braking, the separation between the HeWD and MS star will decrease to the point at which the MS star fills its Roche lobe.
If the MS star has a low mass, $\MMS \le 0.7\,\Msolar$, then it has a substantial convective envelope and mass transfer is expected to be dynamically unstable and lead to a merger if $\MMS / \MWD > 0.695$ \citep{Hurley02,Shen2009}.
It is also possible that orbital decay in a nova common envelope at the onset of mass transfer causes systems with $\MMS / \MWD < 0.695$ to merge \citep{Nelemans2016}.
If the MS star is more massive, $\MMS > 0.7\,\Msolar$, then mass transfer is expected to take place on a thermal time-scale; this may lead to a delayed dynamical instability, and thus also result in a merger \citep{Nelemans2016}.
The immediate products of these HeWD+MS mergers are expected to be red-giant-branch (RGB) like stars \citep{Hurley00,Hurley02,Clausen2011}, however, they may evolve quite differently from normal RGB stars because they form with cool and very degenerate cores.
Some such remnants are expected to ignite helium with a low envelope mass and thus to become hot subdwarfs \citep{Clausen2011}.

In this paper, we compute detailed models of the remnants of HeWD+MS mergers.
We aim to identify the pre-merger conditions that lead to the formation of hot subdwarfs, and to compare to observed stars that may have evolved through this channel.

\section{Methods}
We use the stellar evolution code \texttt{MESA} \citep[Modules for Experiments in Stellar Astrophysics v8118;][]{paxton11,paxton13,paxton15} to model merger remnants.
We start with models of HeWDs onto which we rapidly accrete H-rich matter.
Because it is difficult to control the mass of a HeWD produced in a full binary-star evolution calculation, we adopt an artificial method.
Starting with a $1.5\,\Msolar$ zero-age main-sequence star (metallicity $Z=0.02$), evolution is computed until the He core reaches one of $0.250$, $0.275$, $0.300$, $0.325$, $0.350$, $0.375$, or $0.400\,\Msolar$.
Nucleosynthesis is switched off and a high mass-loss rate is applied to remove the hydrogen envelope completely, leaving a model of an exposed He core.
The model evolves straight to the WD track without He ignition.
Once the logarithmic surface luminosity $\log (L/\Lsolar) = -2$ (see \citet{Zhang12a} for details), we stop the evolution.
This luminosity is chosen to be sufficiently low that stars of all the chosen masses are on the WD track at that stage, but are also not so cool that there are convergence difficulties when accretion is switched on.
This procedure produces models of HeWDs of the required masses which are used in subsequent steps.

\begin{figure}
\plotone{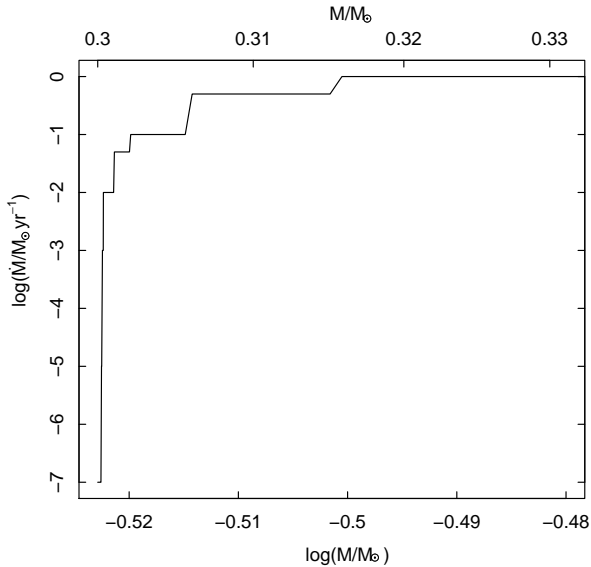}\caption[]{A step-by-step mass accretion process for a $0.300\,\Msolar$ HeWD model.
The rate of accretion of H-rich matter is gradually increased from $10^{-7}\,\Msolar\,\rm{yr}^{-1}$ to $1\,\Msolar\,\rm{yr}^{-1}$.
} \label{1}
\end{figure}

There are currently no numerical simulations of the merger of a HeWD with a low-mass MS star.
We assume that the merger process may take from days to years, a similar time-scale to a CE phase \citep{Ivanova2011,Passy12,Ivanova2013}, and represent the merger by accretion at a rate of $1\,\Msolar\,\rm{yr}^{-1}$ onto a HeWD.
The material accreted from a MS star has He mass fraction $Y=0.28$ and $Z=0.02$ with the scaled metal mixture of \citet{Grevesse1998}.
To converge models accreting at $1\,\Msolar\,\rm{yr}^{-1}$, it is necessary to gradually increase the accretion rate; we increase the rate from $10^{-7}\,\Msolar\,\rm{yr}^{-1}$ to $1\,\Msolar\,\rm{yr}^{-1}$, adjusting the step-by-step increase to ensure convergence (see Fig.~\ref{1} for an example).

Details of the \texttt{MESA} input parameters chosen are given in Appendix~\ref{sec:mesa_inlist}.
The ratio of mixing length to local pressure scale height is set to $\alpha = l/H_{\rm p} = 1.9179$, as found by the solar calibration of \citet{paxton11}.
The opacity tables are from \citet{Iglesias1996} and \citet{Ferguson2005}.
Because the abundances of carbon and oxygen in the interior change after a He flash, we use the OPAL Type 2 opacity tables.
The outer boundary condition is chosen to be an Eddington gray photosphere.
The post-merger models evolve with mass loss when the effective temperature is below $10^4\,\rm{K}$: when the stars have He cores, mass is lost according to Reimers' formula with $\eta_{\rm R} = 0.5$; when the stars have carbon--oxygen cores, mass is lost according to Bl\"ocker's formula with $\eta_{\rm B} = 0.5$ \citep{Bloecker1995, Schindler2015}.
In our models, mixing is due to convection in convective regions and atomic diffusion in radiative regions.
We do not consider overshooting, rotational, semiconvective or thermohaline mixing.
The implementation of atomic diffusion in \texttt{MESA} is based on that of \citet{thoul94}.
Diffusion includes the processes of gravitational settling, thermal diffusion and concentration diffusion.
The atomic diffusion coefficients are those calculated by \citet{paquette86}.
Five species are chosen as representatives for the diffusion calculations: \iso{1}{H}, \iso{4}{He}, \iso{12}{C}, \iso{14}{N}, and \iso{16}{O}.
Nuclear reactions are treated with the `agb.net' network, which includes $17$
nuclides: \iso{1}{H}, \iso{2}{H}, \iso{3}{He}, \iso{4}{He}, \iso{7}{Li}, \iso{7}{Be}, \iso{8}{B}, \iso{12}{C}, \iso{13}{C},
\iso{13}{N}, \iso{14}{N}, \iso{15}{N}, \iso{16}{O}, \iso{17}{O}, \iso{18}{O}, \iso{19}{F} and \iso{22}{Ne}.

\section{Merger remnants}
\begin{figure}
  \plotone{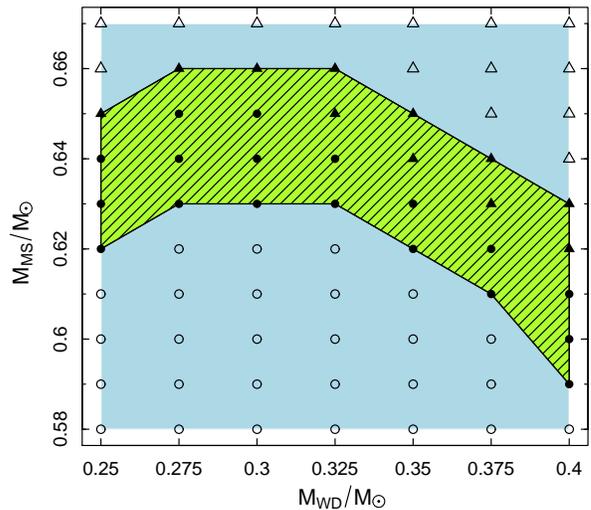}\caption[]{Models of HeWD+MS merger remnants in the $\MWD$--$\MMS$ plane.
    The symbols indicate evolutionary paths: 1.\ HeWDs (open circles), 2.\ Late hot flasher hot subdwarfs (filled circles), 3.\ Early hot flasher hot subdwarfs (solid triangles), 4.\ Normal horizontal branch stars (open triangles).
    The shaded band indicates mergers that result in hot subdwarfs.
} \label{2}
\end{figure}

\begin{figure}
\plotone{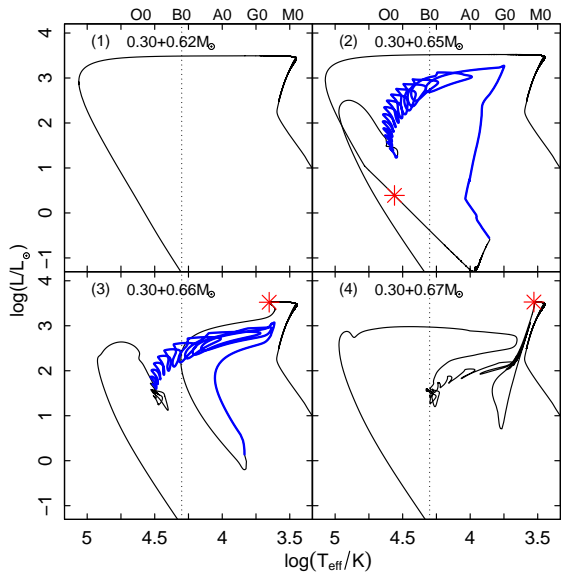}\caption[]{Evolution in the HR diagram of HeWD+MS merger remnants of the given pre-merger $\MWD$+$\MMS$.
A star indicates the first He flash.
Thick lines indicate He-rich phases, $\nHe \ge 0.1$.
Dashed lines indicate the minimum effective temperature required for a star to be a hot subdwarf, $20\,\rm{kK}$.
}\label{3}
\end{figure}

Using the method described, we compute models of $70$ HeWD+MS merger remnants with pre-merger HeWD masses ($\MWD$) from $0.250$ to $0.400\,\Msolar$ in steps of $0.025\,\Msolar$ and MS masses ($\MMS$) from $0.580$ to $0.670\,\Msolar$ in steps of $0.010\,\Msolar$, as shown in Fig.~\ref{2}.
In all cases, an RGB-like star forms, made of a very degenerate He core surrounded by an extended hydrogen envelope.
As in an RGB star, hydrogen burns in a shell, and mass is lost from the surface.
The fresh He produced inside the H-burning shell is added to the He core, compressed and heated.
In the subsequent evolution the star follows one of four distinct paths.
Fig.~\ref{3} shows these different paths in the theoretical Hertzsprung--Russell (HR) diagram, demonstrated by the remnants of the mergers of $0.300\,\Msolar$ HeWDs with $0.620$, $0.650$, $0.660$, and $0.670\,\Msolar$ MS stars.
The paths are as follows:
\begin{enumerate}
\item Do not ignite He, become a HeWD;
\item Ignite He in a late He flash as a HeWD, become a hot subdwarf;
\item Ignite He in an early He flash as a plateau star, become a hot subdwarf;
\item Ignite He and become a normal horizontal branch star, with $T_{\rm{eff}}$ insufficiently high to be a hot subdwarf.
\end{enumerate}
These paths are also those available to stripped RGB stars, as found by previous studies \citep{Castellani1993,dcruz96,sweigart97,Brown2001,Lanz2004,cassisi03,Miller08,lei15}.
These authors found that early hot flashers become extreme horizontal branch stars with hydrogen-rich envelopes unaffected by the He flash, while late hot flashers become He-burning stars with surfaces that may have been affected by mixing during the He flash.
Our merger remnants follow these evolutionary paths without the need for mass-loss enhanced above standard RGB Reimers' rates.

It is the $29$ merger remnants that become hot subdwarfs (paths 2 and 3) on which we focus (Fig.~\ref{2}).
In these merger remnants, once the He core mass grows sufficiently, a He-burning shell is ignited and moves inwards in a series of shell flashes.
The core masses at ignition are larger than the usual value of about $0.47\,\Msolar$ found in normal RGB stars \citep{Han02,Han03} because the total entropy of a HeWD at the moment of merger is much lower than the entropy of a He core at the start of normal red-giant evolution; it thus requires more compression and heating to reach conditions that allow He ignition.

\subsection{Enrichment of helium}
\begin{figure}
\plotone{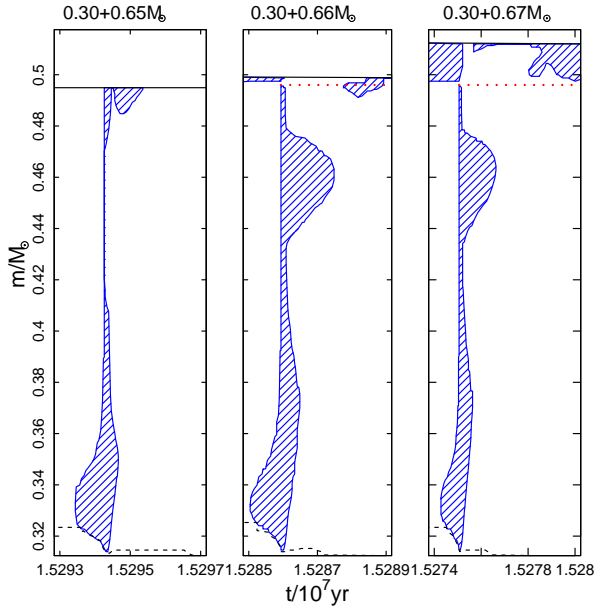}\caption[]{Kippenhahn diagram for a small period of
evolution around the time of the first He flash in $0.300+0.650$, $0.300+0.660$ and $0.300$+$0.670\,\Msolar$ HeWD+MS merger remnants.
By this point in their evolution, the stars have lost most of their envelope in a stellar wind.
Shaded areas indicate convection zones.
Solid lines indicate the stellar surfaces.
Dashed lines indicate the temperature maxima.
Horizontal (red) dotted lines indicate the highest points in the envelope reached by flash-driven convection zones.
} \label{4}
\end{figure}

In the merger remnants that become hot subdwarfs, the first and strongest He flash drives a strong convection zone towards the surface.
The He-burning shell is hot enough ($T > 10^8\,\rm{K}$) for 3$\alpha$ and $\iso{14}{N}(\alpha,\gamma)\iso{18}{O}$ burning, and even for the subsequent $\iso{18}{O}(\alpha,\gamma)\iso{22}{Ne}$ burning.
Depending on the details of ignition, in both early and late hot flashers the products of this He burning may be transferred towards the surface by convection.
For instance, Fig.~\ref{4} shows a Kippenhahn diagram for a small period of
evolution following mergers of $\MWD$+$\MMS=0.300$+$0.650$, $0.300$+$0.660$, and $0.300$+$0.670\,\Msolar$.
By this point in their evolution, the stars have lost most of their envelope in a stellar wind and have masses $0.495$, $0.499$ and $0.512\Msolar$, respectively.
In the $0.300$+$0.650\,\Msolar$ remnant, a late hot flasher, the hydrogen envelope is of very low mass and the first He flash drives a convection zone from the flash zone directly to the surface, yielding a maximum $\nHe=0.995$ ($Y=0.954$).
In the $0.300$+$0.660\,\Msolar$ remnant, an early hot flasher, the upper boundary of convection zone is very close to surface; after the He-shell flash, the hydrogen shell re-ignites, the star expands, and initiates deep opacity-driven surface convection.
This convection zone extends down into processed material from the flash-driven convection zone.
Some of that material, composed mostly of He with some other newly produced elements, is dredged to the surface, yielding a maximum $\nHe=0.331$ ($Y=0.636$).
In the $0.300$+$0.670\,\Msolar$ remnant, an early hot flasher, the He-flash driven and surface-opacity driven convection zones do not touch each other; He-flash produced elements do not appear at the surface and there is no He enrichment.

The evolution of these remnants is similar to that of stripped RGB stars, but there are some differences.
Notably, in early hot flasher merger remnants, He can be mixed to the surface in the first He flash; this is not found in stripped RGB star models \citep{Brown2001}.
In both cases ignition occurs off-center, but in merger remnants the He flash occurs closer to the surface of the star.
This allows convection to reach the surface and dredge up He and He-burning products.
For example, in our $0.300$+$0.660\,\Msolar$ merger remnant, the flash occurs at a Lagrangian mass $m=0.31\,\Msolar$, while in a comparable stripped RGB star the flash occurs at $m=0.18\,\Msolar$ \citep{Brown2001}.
In HeWD+MS merger remnants the He core was previously a HeWD and so is cooler than the core of a normal RGB star.

After the first and strongest He flash, the front of the He-burning flame continues to move inwards and 3$\alpha$ and other $\alpha$-capture reactions heat the compact core, lifting the electron degeneracy therein.
The subsequent He-flash convection zones never reach the surface or mix with surface-opacity driven convection zones.
Thus the surface mass fraction of He depends entirely on the first He flash and envelope mixing afterwards.
After a few Myrs, the He-burning flame reaches the center and true core He burning begins.
The models reach the He-burning main sequence (or zero-age extended horizontal branch) in a few tens of Myrs.

\begin{figure}
\plotone{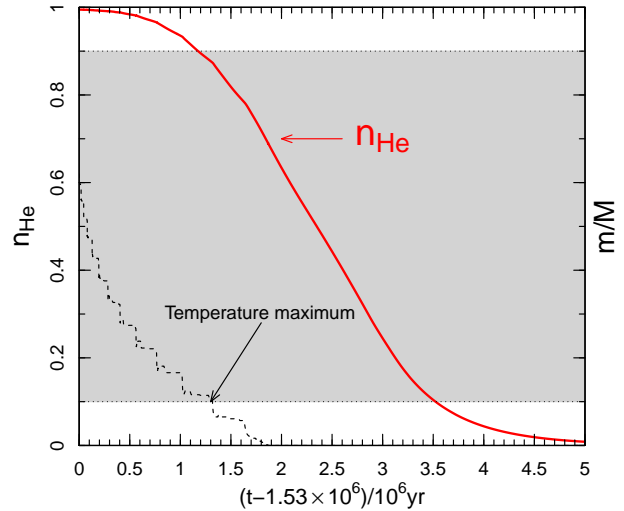}\caption[]{Evolution of the surface He number fraction after the first He flash in a $\MWD$+$\MMS=0.300$+$0.650\,\Msolar$ merger remnant.
The dashed line indicates the location of the temperature maximum.
The number $1.53 \times 10^6\,\rm{yr}$ represents the time at the first He flash.
The shaded region corresponds to the definition of iHe-rich hot subdwarfs ($\nHe = 10$--$90\%$).}
\label{5}
\end{figure}

\subsection{Sinking of helium}
While the He flashes approach the center of the star and the star performs loops in the HR diagram, heavier elements near the surface diffuse downwards and finally produce an almost pure H envelope.
Fig.~\ref{5} shows that it takes about $5\,\rm{Myr}$ for the $0.300$+$0.650\,\Msolar$ remnant to convert the He-rich surface produced at the first He flash into an almost pure H-dominated surface.
The surface abundances of carbon, nitrogen and oxygen also drop rapidly during the same interval.
The diffusion process operates faster than inwards He burning, so the stars already have a H-rich envelope before they become He-burning main sequence hot subdwarfs.
The He-rich phases ($\nHe \ge 0.1$) are indicated by thick lines in Fig.~\ref{3}.
The remnants spend about $70\,\rm{Myr}$ as He-burning H-rich hot subdwarfs.

\subsection{Resolution sensitivity}
\begin{figure}
  \plotone{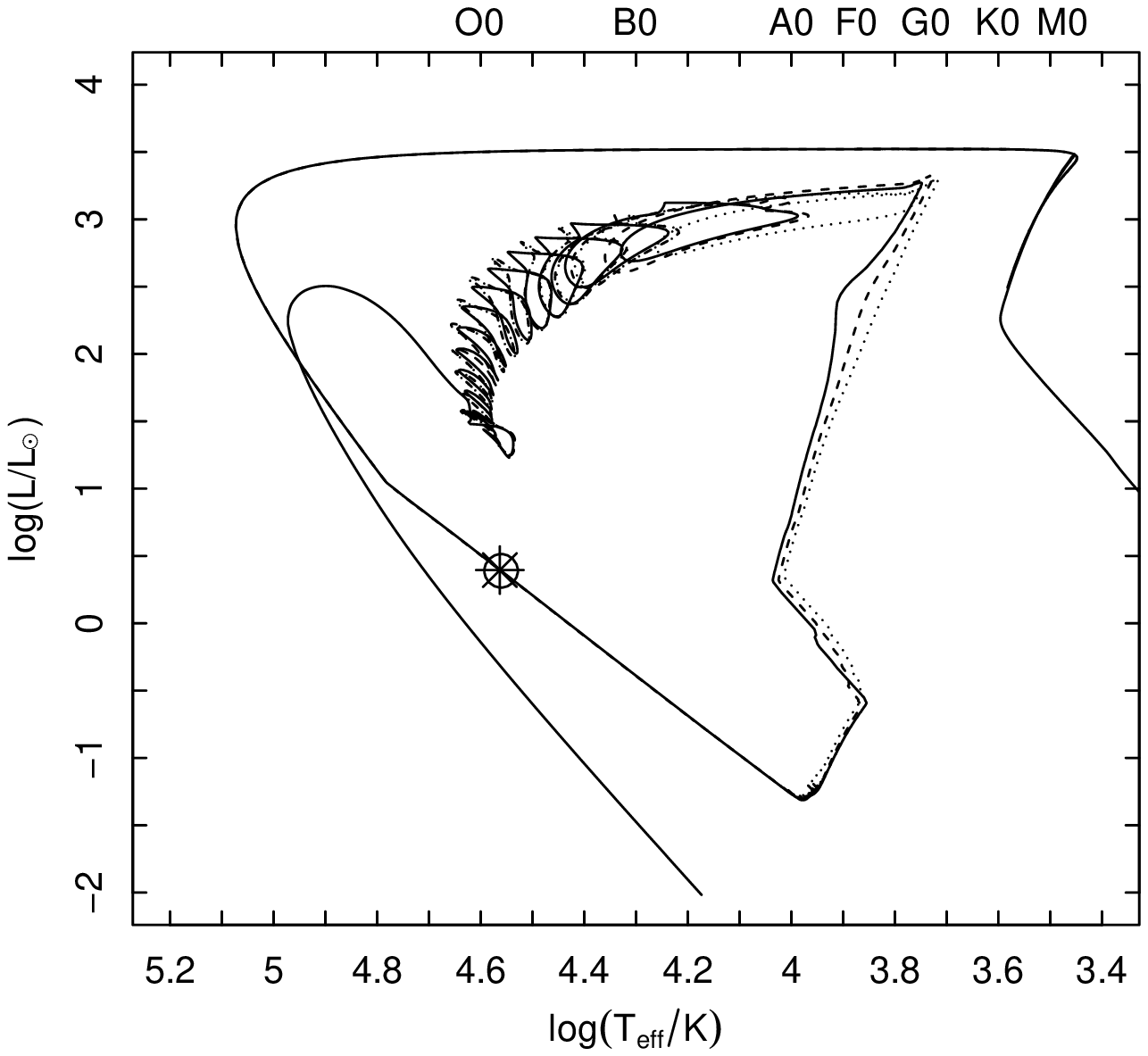}\caption[]{Evolution in the HR diagram after a $0.300$+$0.650$ HeWD+MS merger when different time and space resolution  are used to represent the merger process, i.e., solid line for \texttt{mesh\_delta\_coeff=2} and \texttt{varcontrol\_target=1d-3}; dashed line for \texttt{mesh\_delta\_coeff=1} and \texttt{varcontrol\_target=1d-4}; and dotted line for \texttt{mesh\_delta\_coeff=0.5} and \texttt{varcontrol\_target=1d-4}.}
 \label{6}
\end{figure}

The evolution of hot flasher models can be affected by their time and space resolution.
We check that the $0.300$+$0.650\,\Msolar$ remnant is converged in this sense by computing additional sequences with an increased number of meshpoints and decreased timestep between models, by decreasing \texttt{mesh\_delta\_coeff} from $2$ to $1$ and $0.5$, and decreasing \texttt{varcontrol\_target} from $10^{-3}$ to $10^{-4}$.
Fig.~\ref{6} shows that the evolution in the HR diagram of the three evolutionary sequences is similar.
In all three cases, the He-flash happens at the same stage of evolution, as indicated by the open circle and cross in Fig.~\ref{6}.
The maximum surface He number fractions after the first He flash are also similar: $\nHe=0.995, 0.993, 0.993$ from the standard resolution case to the high resolution case.

\subsection{Accretion rate sensitivity}
\begin{figure}
\plotone{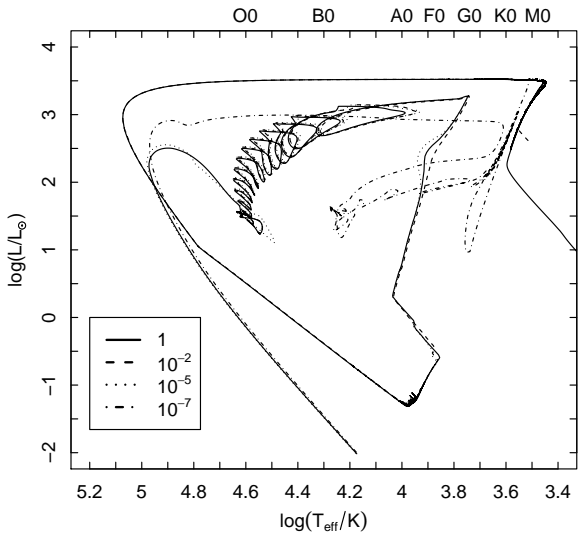}\caption[]{Evolution in the HR diagram after a $0.300$+$0.650$ HeWD+MS merger when different accretion rates are used to represent the merger process, i.e., $1$, $10^{-2}$, $10^{-5}$, $10^{-7}\,\Msolar\,\rm{yr}^{-1}$ model, respectively.
} \label{7}
\end{figure}

In our main set of models, matter is accreted at $1\,\Msolar\,\rm{yr}^{-1}$ to represent the merger process.
To examine the importance of this assumed accretion rate, we compute models of $0.300$+$0.650\,\Msolar$ merger remnants in which mass is accreted at $1$, $10^{-2}$, $10^{-5}$ or $10^{-7}\,\Msolar\,\rm{yr}^{-1}$.
Fig.~\ref{7} shows that the evolution in the HR diagram for the first three cases is very similar at the late stages of interest to this work.
Only when the accretion rate is $10^{-7}\,\Msolar\,\rm{yr}^{-1}$ do we see a difference from our standard model.
This case differs from the others because the accretion rate is sufficiently low that the He core grows by $0.0354\,\Msolar$ during the accretion phase.
The star thus follows a different evolutionary path in the post-merger phase.
The post-merger evolution is independent of the accretion rate if it is sufficiently rapid for the core mass to remain constant during the merger phase.
Indeed, the three highest accretion rate models also show similar maximum surface He number fractions after the first He flash: $\nHe=0.995$, $0.994$ and $0.994$, respectively.

\subsection{Summary}
We have confirmed the suggestion that some HeWD+MS merger remnants can become hot subdwarfs.
We have found the range of pre-merger $\MWD$ and $\MMS$ for which hot subdwarfs are formed.
Additionally, our models show that these stars have surfaces that are He-rich, then iHe-rich and finally H-rich.
A merger remnant that evolves in this way spends about $5\,\rm{Myr}$ as an iHe-rich star and $70\,\rm{Myr}$ as a H-rich hot subdwarf.
Unlike stripped RGB stars that ignite He in early hot flashes, early hot flasher HeWD+MS merger remnants can mix He to the surface and become iHe-rich hot subdwarfs.

\section{Population Synthesis}
\begin{figure}
  \plotone{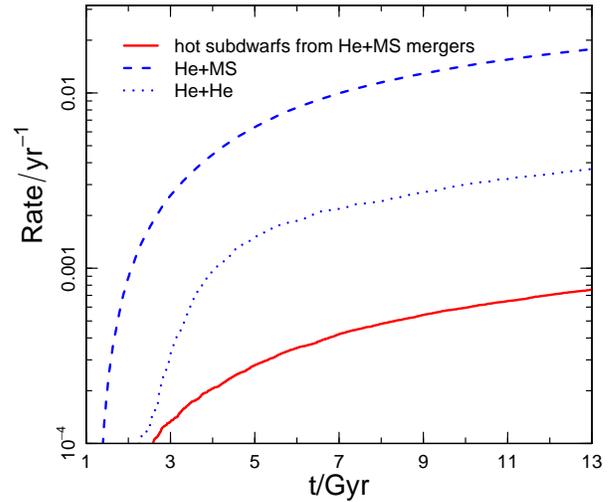}\caption[]{Evolution of the HeWD+MS merger rate (dashed line), the rate of formation of hot subdwarfs from these mergers (solid line) and the HeWD+HeWD merger rate (dotted line) in our model population.
  } \label{8}
\end{figure}

We have found the range of $\MWD$ and $\MMS$ for which HeWD+MS mergers produce hot subdwarfs.
However, if the HeWD+MS merger channel is to make a significant contribution to the hot subdwarf population then the rate of such mergers must be sufficiently high.
One way of checking this is to compare the rate of HeWD+MS mergers that produce hot subdwarfs to the rate of \emph{HeWD+HeWD} mergers.
Double HeWD mergers are widely considered to form isolated hot subdwarfs, thus if the rate of HeWD+MS mergers is comparable then it is reasonable to assume that HeWD+MS merger remnants also contribute to the observed hot subdwarf population.
To estimate the rates of both types of mergers, we compute the properties of a synthetic population of primordial binary systems.
We use a Monte Carlo algorithm to draw $10^7$ sets from our chosen joint distribution of zero-age parameters.
We use a rapid binary evolution code (\texttt{BSE}, \citealt{Hurley00,Hurley02}) to evolve these binary systems for $13\,\rm{Gyr}$ and record the properties of HeWD+MS and HeWD+HeWD mergers.
The parameters in the rapid evolution code in this work are chosen to be the same as those previously used to model the rate of double WD mergers in the Galaxy \citep{Han98, zhang2014}.
Our results are processed to find the properties of a model population with an age of $13\,\rm{Gyr}$ and a constant star formation rate history of $5\,\Msolar\,\rm{yr}^{-1}$, intended to represent the Galaxy \citep{yungelson98}.

For the joint distribution of zero-age parameters, the masses are generated according to the formula of \citet{Eggleton89} and the initial mass function of \citet{Miller79}, with masses in the range $0.08$--$100\,\Msolar$.
The distribution of orbital separations, $p(a)$, is that of \citet{Han98}:
$$ p(a)=\left\{
\begin{array}{lcl}
0.070(a/a_0)^{1.2}      &      & {a \leq a_0}\\
0.070                   &      & {a_0 \le a \le a_1},\\
\end{array} \right. $$
where $a_0=10\,\Rsolar$, $a_1=5.75 \times 10^6\,\Rsolar = 0.13\,\rm{pc}$.

\begin{figure}
  \plotone{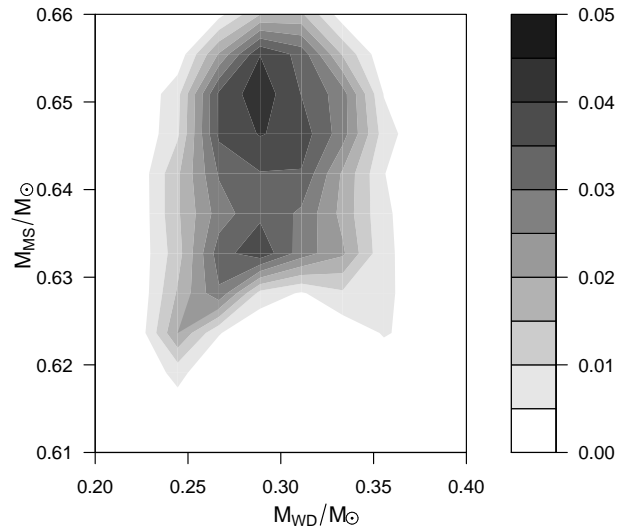}\caption[]{Number fraction of hot subdwarfs formed from HeWD+MS mergers as a function of $\MWD$ and $\MMS$ in the model population.}
 \label{9}
\end{figure}

Of the $10^7$ binary systems, $43\,347$ undergo HeWD+MS mergers.
Not all of these HeWD+MS mergers form a hot subdwarf: to do so requires that the pre-merger masses are in the correct region of Fig.~\ref{2}.
Thus only $1913$ pairs can produce hot subdwarfs.
Fig.~\ref{8} shows the evolution of the HeWD+MS merger rate, the birthrate of hot subdwarfs formed through HeWD+MS mergers and the HeWD+HeWD merger rate.
At $13\,\rm{Gyr}$, hot subdwarfs are formed through HeWD+MS mergers at a rate of $7.57 \times 10^{-4}\,{\rm yr}^{-1}$, about $20\,\%$ of the HeWD+HeWD merger rate of $3.7 \times 10^{-3}\,{\rm yr}^{-1}$, a result that indicates that HeWD+MS mergers may also contribute to the hot subdwarf population.
Fig.~\ref{9} shows the number of hot subdwarfs formed from HeWD+MS mergers as a function of $\MWD$ and $\MMS$ at $13\,\rm{Gyr}$.

\section{Comparison with observation}
\begin{figure}
  \plotone{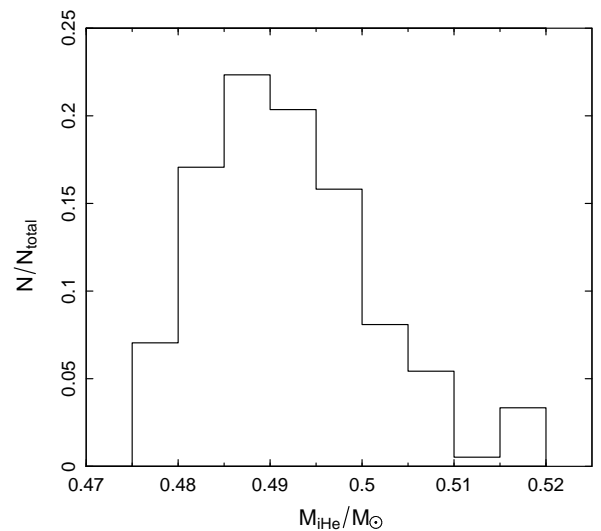}\caption[]{Mass distribution of iHe-rich hot subdwarfs in the model population.} \label{10}
\end{figure}

Having found that model HeWD+MS merger remnants can become iHe-rich and H-rich hot subdwarfs, we compare their properties in more detail to observed examples of such stars.
We use our model population to compute the theoretical distribution of atmospheric parameters for stars formed through this channel.
In the model population, most of the iHe-rich hot subdwarfs formed through the HeWD+MS merger channel have masses in the range $0.48$ to $0.50\,\Msolar$ and a few have masses up to $0.52\,\Msolar$ (Fig.~\ref{10}).

\subsection{Intermediate He-rich hot subdwarfs}
\begin{figure}
\plotone{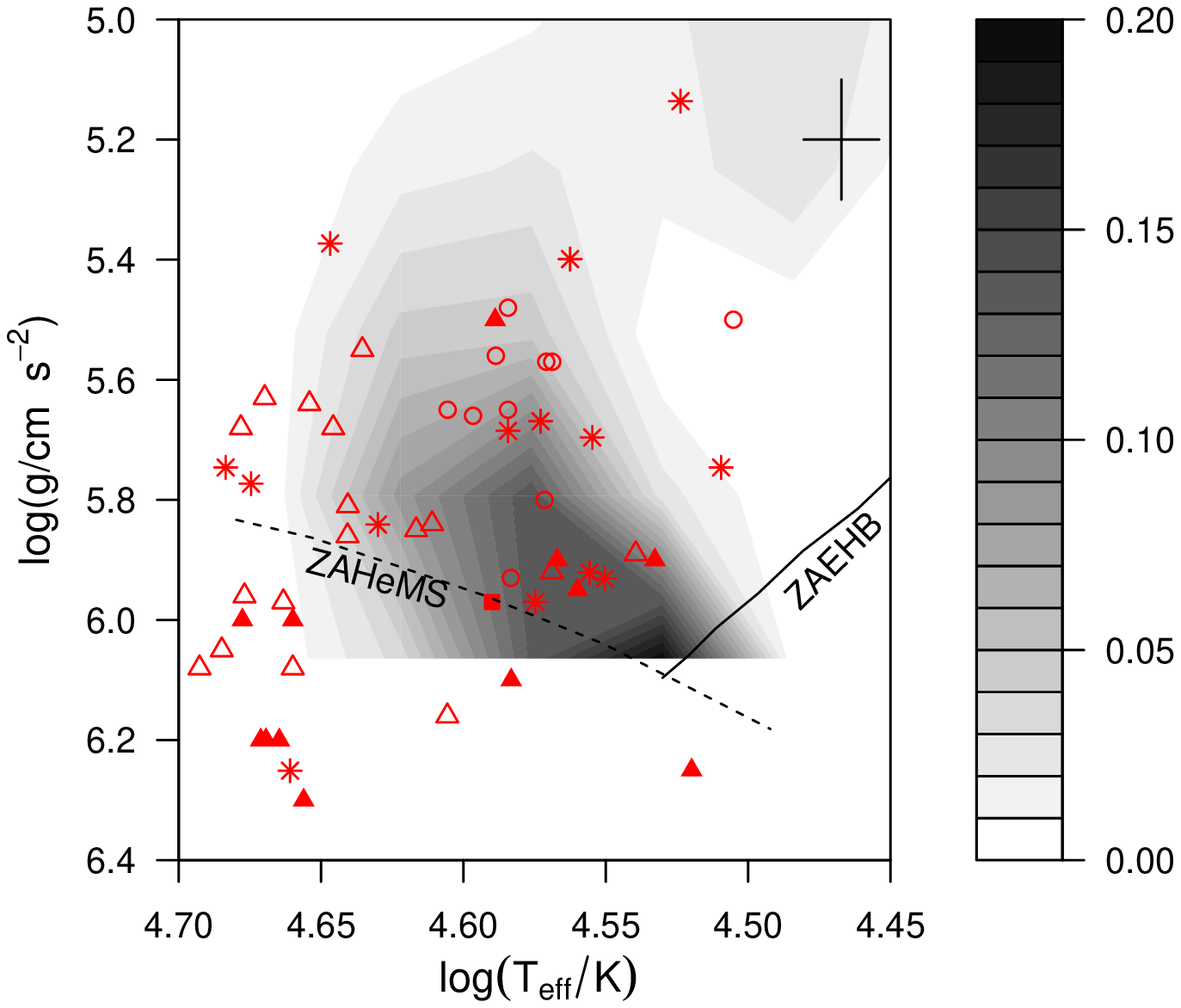}\caption[]{Observed iHe-rich hot subdwarfs in the effective temperature--surface gravity plane.
The theoretical distribution of iHe-rich hot subdwarfs formed through HeWD+MS mergers is indicated by the gray scale.
Symbols represent observed iHe-rich hot subdwarfs  from the samples of \citet{Luo2016} (asterisks), \citet{Nemeth12} (open triangles),  \citet{Naslim10,Naslim11,Naslim13} (open circles), \citet{Jeffery16} (solid squares) and \citet{Edelmann03} (solid triangles).
Average errors are indicated by a cross, upper right.
} \label{11}
\end{figure}

\begin{figure}
  \plotone{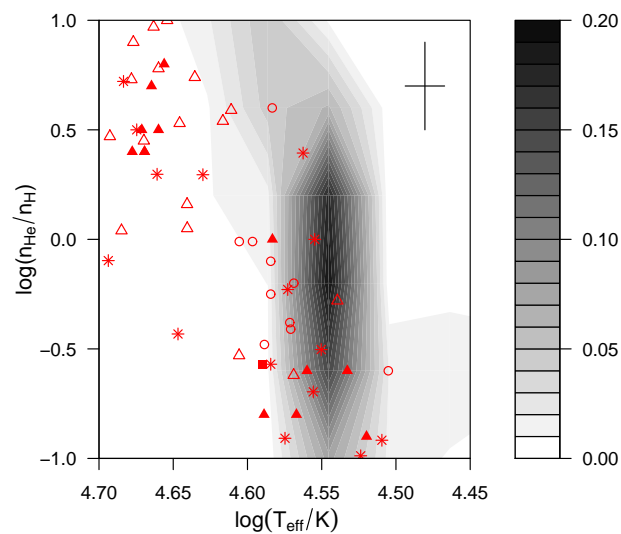}\caption[]{As Fig.\,\ref{11} for the effective temperature--surface He abundance plane.
} \label{12}
\end{figure}

We compile a sample of all known iHe-rich hot subdwarfs \citep[those with $\nHe=10$--$90\%$;][]{Edelmann03,Naslim10,Naslim11,Nemeth12,Naslim13,Jeffery16,Luo2016}.
Fig.~\ref{11} shows, in the $T_{\rm{eff}}$--$\log g$ plane, these stars and the corresponding theoretical distribution of iHe-rich stars formed through HeWD+MS mergers.
The figure shows that the cooler ($30$--$40\,\rm{kK}$) iHe-rich hot subdwarfs are possible HeWD+MS merger remnants, but that not all iHe-rich hot subdwarfs can be explained as having evolved through this channel.
Also, during iHe-rich phases our models have $\log g \la 6.06$, so there is a sharp boundary to the region in which iHe-rich HeWD+MS merger remnants are found.
The observed stars outside this region can also not be explained as HeWD+MS merger remnants.

Fig.~\ref{12} compares models and observation in the $T_{\rm{eff}}$--surface He abundance plane.
This figure again shows that the model HeWD+MS merger remnants have too small a range in these parameters to explain all iHe-rich stars.
Only the cooler and less He-rich stars can be explained as HeWD+MS merger remnants.
Spectroscopically, these would be labeled He-sdOB stars, as distinct from the hotter He-sdO stars.

\subsection{H-rich hot subdwarfs}
\begin{figure}
\plotone{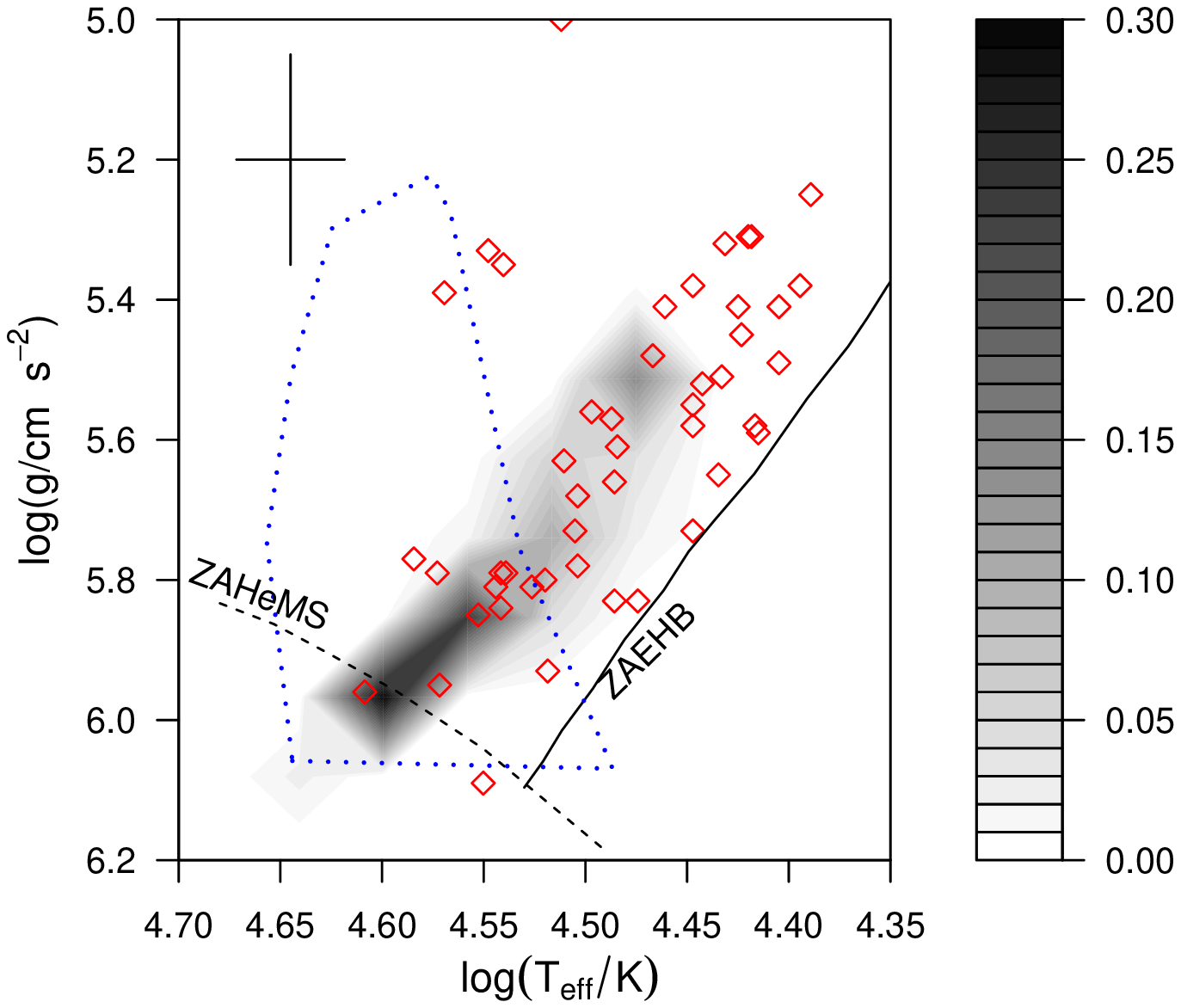}\caption[]{Observed H-rich hot subdwarfs in the effective temperature--surface gravity plane.
The theoretical distribution of H-rich hot subdwarfs formed through HeWD+MS merger remnants is indicated by the gray scale.
Symbols represent observed isolated H-rich hot subdwarfs observed by \citet{Geier13a}, \citet{Geier13b} and \citet{Fontaine2012}.
The dotted line indicates the theoretical iHe-rich hot subdwarf dominated zone from Fig.~\ref{11}.
Average errors are indicated by a cross, upper left.
} \label{13}
\end{figure}

After the iHe-rich phase, the remnants evolve to become H-rich hot subdwarfs.
The duration of this phase is more than $14$ times that of the iHe-rich phase, thus we expect to observe more HeWD+MS merger remnants in the H-rich phase.
As merger remnants, they are all single stars, so we should compare to a sample of isolated H-rich hot subdwarfs.
We compare to the sample of such stars compiled by \citet{Hall2016} from observations by \citet{Geier13a}, \citet{Geier13b} and \citet{Fontaine2012}.
Fig.~\ref{13} shows, in the $T_{\rm{eff}}$--$\log g$ plane, these stars and the corresponding theoretical distribution of H-rich stars formed through HeWD+MS mergers.
The figure shows that some H-rich hot subdwarfs can be explained as HeWD+MS merger remnants.

\section{Discussion and conclusion}
We have modeled the remnants of HeWD+MS mergers.
Such mergers had previously been identified as a channel to the formation of H-rich hot subdwarfs, as we confirm here.
Population synthesis shows that HeWD+MS mergers leading to the formation of hot subdwarfs occur at a rate of about $20\,\%$ of that of HeWD+HeWD mergers.
Additionally, we identify HeWD+MS mergers as one channel to the formation of iHe-rich hot subdwarfs, although not all members of this diverse class can be explained by our models.
In the cases in which an iHe-rich surface is formed after the merger, the process proceeds as follows.
The He core grows by H-shell burning until sufficiently massive for the first and strongest He flash to be ignited and drive a very deep convection zone which reaches the surface.
This convection zone extends down into a region processed during the He flash and some of the matter there, composed mostly of He with some newly produced elements, is dredged to the surface and makes its composition iHe-rich.
Heavier elements sink to leave a H-rich surface as the star approaches the He-burning main sequence.

The surface abundances of helium and other elements are affected by the depth of the first helium flash and the depth of subsequent surface convection zones in HeWD+MS merger remnants.
These features are dominated by the mass of the MS companion,
the merger accretion efficiency, atomic diffusion and the treatment of convective mixing.
A few iHe-rich hot subdwarf stars have extraordinary surface compositions,
with very large surface abundances of lead, zirconium, germanium and yttrium.
These neutron-capture elements may form in the hot helium burning layer of our models or in supernova explosions in their companions,
or be a product of extreme chemical stratification as a result of radiative levitation in the stellar photosphere.

We have only considered HeWD+MS mergers.
It may be that hot subdwarfs -- iHe-rich or otherwise -- can also be formed in the mergers of HeWDs with evolved stars such as Hertzsprung Gap or RGB stars.
The rate of such mergers may exceed HeWD+MS and HeWD+HeWD cases, but it is not obvious which pre-merger configurations would lead to hot subdwarfs.
If the companion to the HeWD is disrupted in the merger then the substantial helium mass in evolved stars may lead to an RGB-like structure in which the envelope has a high helium mass fraction.
Aspects such as this would need to be investigated before we can say if hot subdwarfs are expected from mergers of HeWDs with more evolved stars.
We hope that further work will identify other channels to the formation iHe-rich stars, particularly to explain the hot He-rich subdwarfs that are unlikely to be HeWD+MS merger remnants.

\acknowledgments
This work is supported by the grants 10933002 and 11273007 from the National Natural Science Foundation of China, the
Joint Research Fund in Astronomy (U1631236) under cooperative agreement between the National Natural
Science Foundation of China (NSFC) and Chinese Academy of Sciences (CAS),
the China Postdoctoral Science Foundation, and the Fundamental Research Funds for the Central Universities.
Armagh Observatory is supported by a grant from the Northern Ireland Department for Communities.

\bibliographystyle{apj} 
\bibliography{mybib} 

\appendix
\section{\textsc{mesa} inlist}\label{sec:mesa_inlist}
To evolve merger remnants with \textsc{mesa} the parameters that differ from the defaults are as follows:
{\small
\begin{verbatim}

&star_job
   change_net = .true.
   new_net_name = 'agb.net'
/

&controls
   use_Type2_opacities = .true.
   initial_z = 0.02
   Zbase = 0.02

   mixing_length_alpha = 1.9179

   which_atm_option = 'Eddington_grey'

   cool_wind_RGB_scheme = 'Reimers'
   Reimers_scaling_factor = 0.5
   cool_wind_AGB_scheme = 'Blocker'
   Blocker_scaling_factor = 0.5
   RGB_to_AGB_wind_switch = 1d-4

   varcontrol_target = 1d-3
   mesh_delta_coeff = 2

   do_element_diffusion = .true.
   diffusion_dt_limit = 3.15d7
   diffusion_min_dq_at_surface = 1d-12
   surface_avg_abundance_dq = 1d-12
/
\end{verbatim}
}

\end{document}